\documentclass[preprint,showpacs]{revtex4}
\usepackage{graphicx}
\usepackage{amsmath}
\usepackage{amssymb}
\usepackage{color}
\usepackage[utf8]{inputenc}
\usepackage{verbatim}

\usepackage[tight,sf]{subfigure}%

\def\vec #1{{\bf #1}}

\newcommand{\romd}{{\text{d}}}

\begin{document}
\title{Nonlinear buckling and symmetry breaking of a soft elastic sheet sliding on a cylindrical substrate}

\author{Norbert Stoop$^1$}
\author{Martin Michael Müller$^{2,3}$} 

\affiliation{
$^1$ Department of Mathematics, Massachusetts Institute of Technology; 77 Massachusetts Avenue, Cambridge, MA 02139-4307, USA\\
$^2$ Equipe BioPhysStat, LCP-A2MC, Universit\'{e} de Lorraine; 
1 boulevard Arago, 57070 Metz,  France\\
$^3$ Institut Charles Sadron, CNRS-UdS; 23 rue du Loess, BP 84047, 67034 Strasbourg cedex 2, France
}

\date{\today}
   
\begin{abstract}
We consider the axial compression of a thin sheet wrapped around a rigid cylindrical substrate. In contrast to the wrinkling-to-fold transitions exhibited in similar systems, we find that the sheet always buckles into a single symmetric fold, while periodic solutions are unstable. Upon further compression, the solution breaks symmetry and stabilizes into a recumbent fold. Using linear analysis and numerics, we theoretically predict the buckling force and energy as a function of the compressive displacement. We compare our theory to experiments employing cylindrical neoprene sheets and find remarkably good agreement. 
\end{abstract}
 

\maketitle


\section{Introduction}
When you roll up your sleeves to get some work done, you will not pay attention to the intricate folding patterns which form around your arms. 
However, these patterns are not only interesting for graphics designers but of eminent importance for biology and technology alike \cite{Chen10,Li12}: one can find them in the twinkling of an eye \cite{ZhuChen13} as well as in the development of organs such as the brain \cite{vanEssen97,Richman75}, the intestine \cite{CiarlettaBenAmar12A,CiarlettaBenAmar12B}, or the kidney \cite{Kuure00}. Technological applications include structures for optics \cite{Wang11} or microfluidics \cite{Efimenko05} to name just a few. 

One common theme in these examples is that they consist of coupled layers which undergo morphological changes in response to external or internal constraints such as a simple compression or volumetric growth. The materials involved range from swelling hydrogels \cite{BenAmarCiarletta10,Dervaux11} to supported graphene \cite{Arroyo13,Arroyo14} and many others. 
A well-studied setup in this context consists of a stiff membrane attached to a flat elastic or fluid bulk material \cite{Pocivavsek08,Brau11,Brau13}. The interplay between the bending of the sheet and the response of the bulk leads to the formation of wrinkles when the sheet is compressed uniaxially. Beyond a critical compression the wrinkles vanish and localized folds appear in the sheet. Interestingly, the shape of the sheet on the fluid can be found analytically \cite{DiamantWitten11,Rivetti13,DiamantWitten13} as long as the sheet does not touch itself \cite{Santangelo14}. 

When the substrate is not flat, translational invariance is broken and a whole plethora of folding patterns can be found \cite{Chen10,Li12,BenAmarGoriely05,PascalisNapoliTurzi14}. In this article we study a particular type of such a system in a cylindrical geometry. An elastic cylindrical membrane is wrapped around a solid cylinder of same radius and compressed parallel to the axis of symmetry. This simple system is of potential relevance for situations as diverse as intestinal inversion \cite{Daneman96}, the folding of your sleeve (see Fig.~1), or even the neck of hidden-necked turtles \cite{ZhuChen13}. In contrast to the aforementioned flat system we allow the membrane to stretch azimuthally to accommodate to the external stress. Without the solid cylinder constraint the membrane would behave like a hollow cylindrical shell whose mechanics has been studied extensively in the literature \cite{Hoff66,Hunt00}: when compressed the shell develops regular patterns, such as periodic, axisymmetric undulations \cite{Timoshenko59} or the trapezoidal patterns found by Yoshimura in the 1950s \cite{Yoshimura55}. As is easily confirmed by compressing an empty can of soda, axisymmetric modes of deformation are typically unstable. As we will see below, the behavior becomes fundamentally different when the shell enwraps a solid cylinder of same size.

Similar to a ruck in a rug \cite{Domokos03,Vella09,Kolinski09,WagnerVella13} we will consider the case in which the sheet can slide on the cylinder. The only coupling between the sheet and the substrate is \textit{via} the hard cylinder constraint. There is no elastic response between the two as is typically the case for cylindrical core-shell materials \cite{Patricio14}. To simplify the theoretical treatment we assume that the sheet is unstretchable in the axial direction. This assumption will be validated by experiments with neoprene sheets and finite element simulations. 

We start with the presentation of our model in Sec.~\ref{sec:model}. The resulting shape equations are axisymmetric and can be linearized  and solved in lowest order of the axial compression as shown in Sec.~\ref{sec:linearization}. In Sec.~\ref{sec:numericalsolutionsexperiments} solutions of the full nonlinear system are found numerically and compared to experiments with neoprene sheets.


\section{Model\label{sec:model}}

\begin{figure}[t]

\begin{center}
\subfigure[][]{\label{fig:pullover}\includegraphics[height=0.3\textwidth]{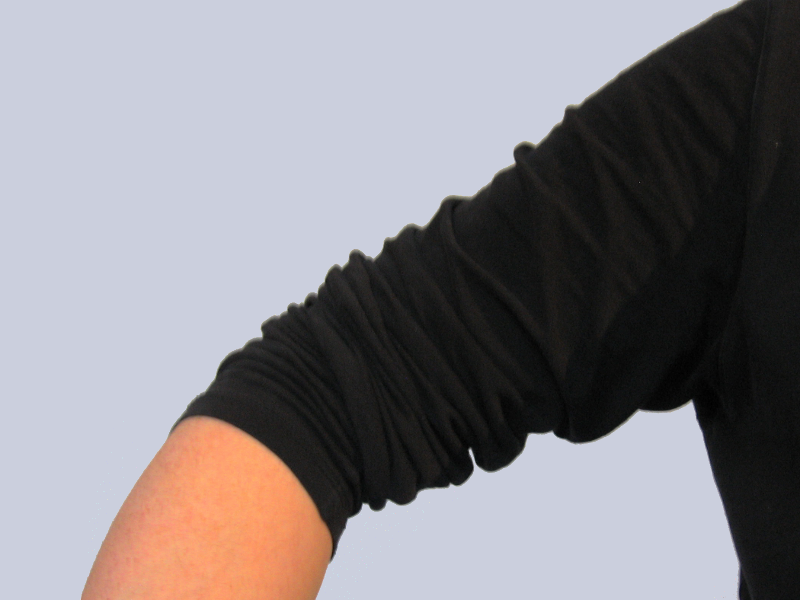}}
\\
\subfigure[][]{\label{fig:system1}\includegraphics[scale=0.31]{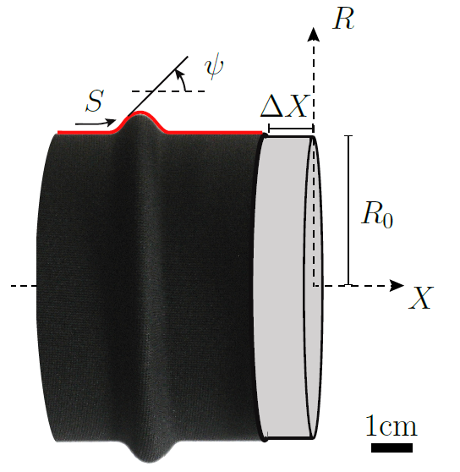}}
\hspace{0.25cm}
\raisebox{2.cm}{$\Rightarrow$}
\hspace{0.5cm}
\subfigure[][]{\label{fig:system2}\includegraphics[scale=0.28]{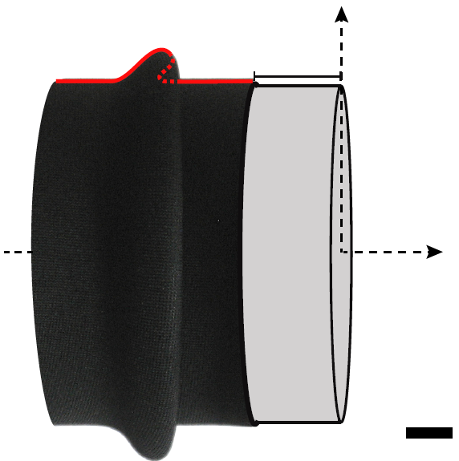}}
\end{center}
\caption{\textit{Top}: 
(a) Rolling up the sleeve. \textit{Bottom}: The studied system (b) before and (c) after the tipping point (overlay of a photograph of the neoprene sheet used in the experiments with a sketch of the variables of the theory).}
\end{figure}

We consider a cylindrical elastic sheet of thickness $h$ which enwraps a cylinder of radius $R_0$ (see Fig.~\ref{fig:system1}). The axis of symmetry is oriented along the $X$ axis whereas we use $R$ as the variable for the radial displacement. 
When the sheet is compressed with a fixed displacement $\Delta X$, it buckles out of its reference configuration and forms a fold which tips over for large compressions (see Fig.~\ref{fig:system2}). In the following we will use the angle-arc length parametrization which describes the shape of the axisymmetric sheet with the help of the tangent angle $\psi$ as a function of arc length $S$. 

The total elastic energy is given by the sum of stretching and bending contributions. We use the linear-strain model described in the appendix with the elastic energy 
\begin{eqnarray}
E =  \frac{Y h \pi}{1-\nu^2} \int_{0}^{S_c}  \Bigg\{ \frac{(R - R_0)^2}{R_0^2}  +  \frac{h^2}{12} \!\! 
&& \left[ \psi'^2 + \left( \frac{\cos{\psi}}{R_0} - \frac{1}{R_0}\right)^2  \right.
\nonumber \\
&& + \; \left. 2 \nu \psi' \left( \frac{\cos{\psi}}{R_0} - \frac{1}{R_0}\right)  \right]
\Bigg\} \; R_0 \, \romd S
\; .
\label{eq:energy1}
\end{eqnarray}
We are interested in the case where $R_0/h \gg 1$. The stretching term $\frac{h(R - R_0)^2}{R_0^2}\cdot R_0$ clearly diverges as $h/R_0$ approaches $0$, 
and similarly does the first bending contribution from $\psi'^2$. The second bending term $\propto h^3 R_0 (\frac{\cos{\psi}}{R_0} - \frac{1}{R_0} )^2$, however, 
vanishes for large $h/R_0$ since $\cos{\psi}$ is bounded. The last term, which looks like a Gaussian curvature, but is not due to the integration over the 
reference configuration, can be integrated to a term proportional to 
$-\sin{\psi} + \psi$
and is constant due to the fixed angle $\psi=0$ at the boundaries for all configurations. Therefore, we do not need to take it into account either. 
This is true even for small $R_0/h$: Using linear strains, the problem is independent of the material's Poisson ratio. We are left with the simplified model
\begin{equation}
E= \frac{Y h \pi}{1-\nu^2} \int_{0}^{S_c} \left[ \frac{(R - R_0)^2}{R_0^2} + \frac{h^2}{12}\psi'^2 \right] R_0 \, \romd S
\; .
\label{eq:energysimplemodel}
\end{equation}
We define $\kappa = \frac{Y h^3}{12(1-\nu^2)}$ and introduce some additional variable rescaling
\begin{equation}
  s:= \frac{S}{\sqrt{h R_0}} \, , \quad x := \frac{X}{\sqrt{h R_0}} \, , \quad \rho := \frac{R - R_0}{\sqrt{h R_0}} \, 
\end{equation}
in order to write the energy functional as 
\begin{eqnarray}
  e & := & \frac{E}{2\pi R_0 \kappa} = \int_{0}^{s_c} \romd s \; \mathcal{L} 
  \nonumber \\
  & = & \int_{0}^{s_c} \romd s \;  \left[\frac{1}{2} ( \psi'^2  + 12 \rho^2)
    + \lambda_{\rho} \left(\rho' -  \sin{\psi}\right) + \lambda_{x} \left(x' -  \cos{\psi}\right) \right]
  \; ,
  \label{eq:Hamiltonian_1D}
\end{eqnarray}
where the dash denotes derivatives with respect to $s$. The Lagrange multiplier functions  $\lambda_{x}$ and $\lambda_{\rho}$ couple the Cartesian coordinates $x$ and $\rho$ to $\psi$. 
The conjugate momenta are
\begin{equation}
  p_{\psi} = \frac{\partial \mathcal{L}}{\partial \psi'} 
    = \psi' \; ,  \qquad
  p_x = \lambda_x \; ,  \qquad
  p_\rho = \lambda_\rho 
      \label{eq:conjugatemomenta}
  \; .
\end{equation}
We switch to a Hamiltonian formulation with the Hamiltonian
\begin{eqnarray}
  \mathcal{H} & = & \psi' p_{\psi}  + x' p_x + \rho' p_\rho - \mathcal{L}
  \nonumber \\
  & = & \frac{p_\psi^2}{2} -  6  \rho^2 + p_\rho \sin{\psi}+ p_x \cos{\psi}
  \; ,
  \label{eq:Hamiltonian}
\end{eqnarray}
from which we obtain the Hamilton equations:
\begin{subequations}\label{eq:Hamiltonequations}
\begin{eqnarray}
  \psi' & = & \frac{\partial \mathcal{H}}{\partial p_{\psi}} 
    = p_\psi \; ,
  \label{eq:Hamiltonequation1}
  \\
  x' & = & \frac{\partial \mathcal{H}}{\partial p_{x}}  
    = \cos{\psi} \; ,
  \label{eq:Hamiltonequation2}
    \\
  \rho' & = & \frac{\partial \mathcal{H}}{\partial p_{\rho}}  
    = \sin{\psi} \; ,
  \label{eq:Hamiltonequation3}
  \end{eqnarray}
\begin{eqnarray}
  p'_{\psi} & = & -\frac{\partial \mathcal{H}}{\partial \psi} 
    =p_x \sin{\psi} - p_\rho \cos{\psi}  \; ,
  \label{eq:Hamiltonequation4}
  \\
    p'_{x} & = & -\frac{\partial \mathcal{H}}{\partial x} 
    =  0 \; ,
  \label{eq:Hamiltonequation5}
  \\
  p'_{\rho} & = & -\frac{\partial \mathcal{H}}{\partial \rho} 
    =  12 \rho
  \label{eq:Hamiltonequation6}
  \; . 
\end{eqnarray}
\end{subequations}
This set of equations strongly resembles the geometrically nonlinear Euler beam model, which reads \cite{AudolyPomeau}
\begin{equation*}
\psi'' = \bar{F} \sin{\psi} -\bar{F}_n \cos{\psi}\; .
\end{equation*} 
Taking the derivative of Eq.~(\ref{eq:Hamiltonequation1}) and combining it with Eq.~(\ref{eq:Hamiltonequation4}), we get a `modified' Euler beam with
\begin{equation*}
\psi'' = p_x \sin{\psi} - p_\rho \cos{\psi}\; 
\end{equation*} 
in scaled variables. The essential difference is that the resulting 'normal' force component $p_\rho$ is now a function of $s$. 
The Lagrange multiplier $p_x$ is a constant along the contour due to Eq.~(\ref{eq:Hamiltonequation5}) and is directly related to the external horizontal force $F$: 
\begin{equation}
p_x = -\frac{F}{2\pi R \kappa}\sqrt{hR} =: - f
\; ,
\end{equation}
where $f$ is positive when the sheet is compressed. 

To find the shape of a single fold, the Hamilton equations~(\ref{eq:Hamiltonequations}) have to be solved with the appropriate boundary conditions: 
\begin{subequations}  \label{eq:boundaryconditions}
\begin{eqnarray}
  \rho (0) & = & \rho (s_c) = 0  \; ,
  \label{eq:BC1}
  \\
  \psi (0) & = & \psi (s_c) = 0  \; ,
  \label{eq:BC2}
  \\
  \psi'(0) & = & \psi' (s_c) = 0 \; ,
  \label{eq:BC3}
  \\
  \mathcal{H} & = & -f
  \label{eq:BC4}
  \; .
\end{eqnarray}
\end{subequations}
Eq.~(\ref{eq:BC2}) takes into account that the membrane must not have kinks:  
at the contact point $s=0$  the membrane detaches from the cylinder and $\psi$ equals 
$0$; at $s=s_c$, we have $\rho (s_c)=0$ and the profile is horizontal again. 
The contact curvature condition~(\ref{eq:BC3}) results from an energy balance at the contact point \cite{Seifert90,boundcond}, 
whereas the Hamiltonian~(\ref{eq:BC4}) is a constant due to the fact that we do not fix the total arc length $s_c$.


\section{Linearization\label{sec:linearization}}

Owing to the strong nonlinearities, we are unaware of any analytical, closed-form solution of Eqns.~(\ref{eq:Hamiltonequations}). Nonetheless, certain unknowns such as the buckling force and the fold length can already be obtained in good approximation using linearized equations. To derive them, we start from Eqns.~(\ref{eq:Hamiltonequations}) of a single fold 
and eliminate $\rho$ by taking the derivative of Eq.~(\ref{eq:Hamiltonequation6}). This yields 
\begin{align}
p_{\rho}'' &= 12 \sin \psi \; , \\
\psi'' &=  - f \sin\psi - p_{\rho}\cos\psi
\; .
\end{align}
We define 
\begin{equation}
\epsilon = \left(\frac{\Delta x}{s_c}  \right)^{1/2}
\end{equation}
and reparametrize the equations above using $\tilde{s} :=  2s/s_c-1$ to obtain    
\begin{align}
p_{\rho}'' &= 3 s_c^{2} \sin \psi \; , 
\label{eq:prho}\\
\psi'' &=  - \tilde{f} \sin\psi - \frac{s_c^2}{4} p_{\rho} \cos\psi \; , 
\label{eq:ddotpsi}\\
\frac{1}{2} \int_{-1}^{1} \cos \psi \,  \romd \tilde{s} &= 1-\epsilon^2 
\; ,
\label{eq:displacementconstraint}
\end{align}
where $\tilde{f} = f  s_c^2/4$ is positive since the shell is axially compressed. Dashes denote derivatives with respect to $\tilde{s}$ from now on. 
We are interested in solutions where $\epsilon\ll 1$. Therefore, we use the following expansion for $p_{\rho}$, $\psi$, $F$ and $s_c$:
\begin{align}
\psi &= \psi^{(1)} + \psi^{(3)} + \mathcal{O}(\epsilon^5) \; ,\\
p_{\rho} &= p_{\rho}^{(1)} + p_{\rho}^{(3)} + \mathcal{O}(\epsilon^5) \; ,\\
\tilde{f} &=\tilde{f}^{(0)} + \tilde{f}^{(2)} + \mathcal{O}(\epsilon^4) \; ,\\
s_c &=s_c^{(0)} + s_c^{(2)} + \mathcal{O}(\epsilon^4) \; .
\end{align}

Plugging the expansions into Eqns.~(\ref{eq:prho})/(\ref{eq:ddotpsi}) and collecting terms of linear order in $\epsilon$ gives
\begin{align}
p_{\rho}^{(1)''} &= 3 (s_c^{(0)})^2 \psi^{(1)} \; ,\\
\psi^{(1)''} &= -\tilde{f}^{(0)} \psi^{(1)} - \frac{(s_c^{(0)})^2}{4} p_{\rho}^{(1)}
\; .
\end{align}
Combining both equations into one 4th order equation, we finally get
\begin{equation}
\psi^{(1)''''}  + \tilde{f}^{(0)} \psi^{(1)''} + \frac{3 (s_c^{(0)})^4}{4} \psi^{(1)} = 0
\end{equation}
the general solution of which is given by
\begin{eqnarray}
\psi^{(1)}_\text{gen}(\tilde{s}) &=& c_1 e^{\sqrt{\lambda_+^2} \tilde{s}} + c_2 e^{- \sqrt{\lambda_+^2} \tilde{s}} + c_3 e^{\sqrt{\lambda_-^2} \tilde{s}} + c_4 e^{- \sqrt{\lambda_-^2} \tilde{s}}\\
\mbox{with }\; \lambda_\pm^2  &=& \frac{-\tilde{f}^{(0)} \pm \sqrt{(\tilde{f}^{(0)})^2 - 3 (s_c^{(0)})^4}}{2}
\; .
\end{eqnarray}
Note that $\sqrt{\lambda_\pm^2}$ is imaginary since $\tilde{f}^{(0)}>0$. One directly obtains: 
\begin{equation}
   \tilde{f}^{(0)} = |\lambda_+^2| + |\lambda_-^2| \qquad \text{and} \qquad s_c^{(0)} = 2 \sqrt[4]{|\lambda_+^2| |\lambda_-^2|/12}
   \; .
   \label{eq:F0sc}
\end{equation}
For small to moderate displacements, we expect symmetric folds. We thus assume $\psi$ to be asymmetric in $\tilde{s}\rightarrow -\tilde{s}$ and construct asymmetric solutions $\psi^{(1)}(\tilde{s}) = \frac{1}{2} \left( \psi^{(1)}_\text{gen}(\tilde{s}) - \psi^{(1)}_\text{gen}(-\tilde{s})\right)$, leading to
\begin{equation}
\psi^{(1)}(\tilde{s}) = a_1 \sin\left(\sqrt{|\lambda_+^2|} \tilde{s}\right) + a_2 \sin\left( \sqrt{|\lambda_-^2|} \tilde{s} \right)
\; ,
\end{equation}
where the coefficients $a_1=i \, (c_1-c_2)$ and $a_2 =i \, (c_3-c_4)$ are real. 
By restricting ourselves to these solutions, it suffices to satisfy boundary conditions at one side.
Enforcing $\psi^{(1)}(1)=0$, we can determine $a_1$ as
\begin{equation}
a_1 = -a_2 \frac{ \sin\left(\sqrt{|\lambda_-^2|}\right)}{\sin\left(\sqrt{|\lambda_+^2|}\right)}
\end{equation}
leading to
\begin{equation}\label{eq:COND0}
\psi^{(1)}(\tilde{s}) = a_2 \left[-\mbox{csc}\left(\sqrt{|\lambda_+^2|}\right) \sin\left(\sqrt{|\lambda_-^2|}\right)
  \sin\left( \sqrt{|\lambda_+^2|}  \tilde{s} \right) 
 + 
   \sin\left(\sqrt{|\lambda_-^2|} \tilde{s} \right) \right]
   \; .
\end{equation}
We first consider solutions where the curvature $\psi'^{(1)} = 0$ at $\tilde{s}=1$. This leads to the condition
\begin{equation}\label{eq:COND1}
   \sqrt{|\lambda_+^2|} \, \cot{\left(\sqrt{|\lambda_+^2|}\right)} - \sqrt{|\lambda_-^2|} \, \cot{\left(\sqrt{|\lambda_-^2|}\right)} = 0 
   \; .
\end{equation}
We recall that the Hamiltonian of the folded sheet is conserved along $\tilde{s}$. This provides an additional equation to determine admissible values of $\sqrt{|\lambda_\pm^2|}$. In rescaled variables Eq.~(\ref{eq:Hamiltonian}) reads
\begin{equation}\label{eq:rescaledHamiltonian}
\mathcal{H} =  \frac{2 \psi'^2}{s_c^2} - 6 \rho^2 + p_{\rho} \sin \psi - \frac{4 \tilde{f}}{s_c^2} \cos \psi
\; ,
\end{equation}
where $\rho$ is the solution to $\rho' = \frac{s_c}{2} \sin \psi$. From the boundary condition~(\ref{eq:BC3}) we find $\mathcal{H}=-\frac{4 \tilde{f}}{s_c^2}$ for $-1<\tilde{s}<1$. In particular in the middle of the fold ($\tilde{s}=0$) we thus obtain for each order in $\epsilon$
\begin{equation}\label{eq:COND2}
\frac{2 \psi'(0)^2}{s_c^2} - \frac{3 s_c^2}{2} \left( \int_{-1}^0 \sin \psi \, \romd \tilde{s}\right)^2 = 0
\; .
\end{equation}
%
\begin{figure}[t]
\begin{center}
\includegraphics[scale=0.6]{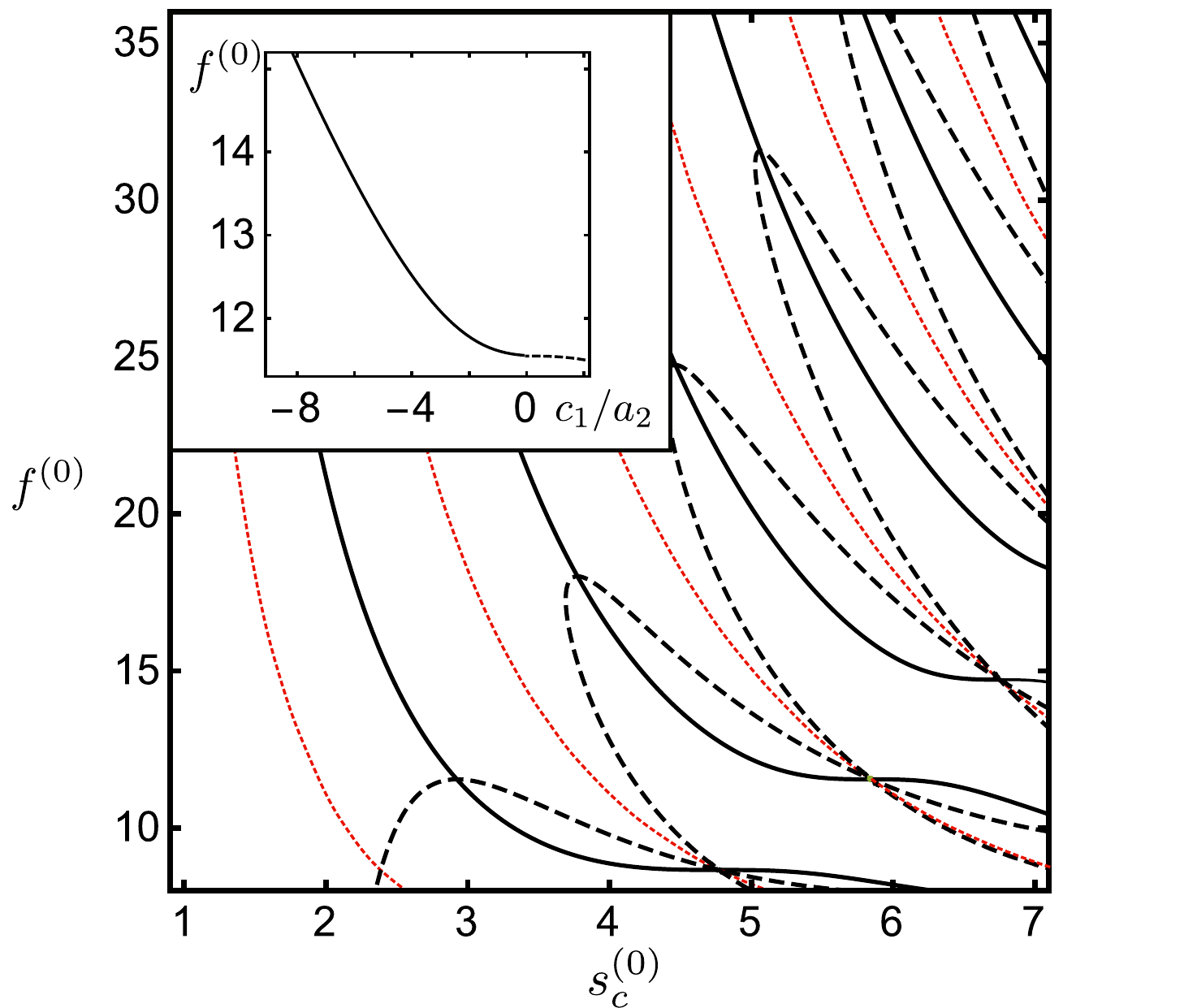}
\end{center}
\caption{Roots of $\mathcal{H}+f^{(0)}$ (dashed) and $\psi^{(1)'}(1)$ (solid black) as a function of $s_c^{(0)}$ and $f^{(0)}$. Crossings between both roots correspond to nontrivial solutions of the linearized fold equation. Dotted red lines denote singularities of $\psi^{(1)'}(1)$. The solution with the lowest force is found for $(s_c^{(0)},f^{(0)}) \approx (2.92, 11.55)$. The inset shows the buckling force as a function of the rescaled curvature boundary condition $\psi^{(1)'}(1) = c_1/a_2$. The force reaches a minimum for $c_1=0$. The dotted line shows the buckling force for unphysical positive values of $c_1/a_2$, where the fold would buckle inwards.
}\label{fig:linearization}
\end{figure}
Fig. \ref{fig:linearization} shows the roots of Eq. (\ref{eq:COND1}) (solid black) and (\ref{eq:COND2}) (dashed black) as a function of $s_c^{(0)}$ and the buckling force $f^{(0)}$. Crossing points correspond to solutions satisfying both conditions. The dotted red lines in Fig. \ref{fig:linearization} denote singularities of Eq. (\ref{eq:COND1}), which excludes, e.g., the point $(s_c^{(0)}, f^{(0)}) \approx (4.8, 8)$ from the set of solutions. Upon inspection, we observe that solutions are found for $\sqrt{|\lambda_+^2|}=\frac{n \pi}{2}$ and  $\sqrt{|\lambda_-^2|}=\frac{m \pi}{2}$ where $m>n>0$ are impair integers. According to Eq.~(\ref{eq:F0sc}), the choice $m=3$ and $n=1$ yields the lowest possible buckling force and thus corresponds to the physical solution with
\begin{equation}\label{eq:lowestsolution}
\tilde{f}^{(0)} = \frac{5\pi^2}{2} \approx 24.67,\;\;\; s_c^{(0)} = \left(\frac{3}{4}\right)^{1/4} \pi \approx 2.92 
\end{equation}
or $f^{(0)}=20/\sqrt{3}\approx 11.55$. The angle $\psi^{(1)}$ at linear order follows as
\begin{equation}
  \psi^{(1)} (\tilde{s}) = a_2 \left(\sin{\frac{\pi}{2}\tilde{s}} + \sin{\frac{3 \pi}{2}\tilde{s}} \right)
  \; .
\end{equation}
It should be noted that the amplitude $a_2$ is still undetermined at this point. It can be found by prescribing the displacement $\epsilon^2$: Integrating the displacement constraint equation Eq.~(\ref{eq:displacementconstraint}) and retaining terms up to order $\epsilon^2$ yields
\begin{equation}\label{eq:amplitude}
1-\epsilon^2 \approx \frac{1}{2}\int_{-1}^{1} \left[ 1- \frac{ \left( \psi^{(1)} \right)^2}{2} \right] \romd \tilde{s}  = 1 - \frac{a_2^2}{2}
\; ,
\end{equation}
and thus $a_2 = - \sqrt{2} \epsilon$, where we chose the negative root in order to obtain positive radial displacements (outward buckling). 
We also note that the maximum radial displacement $\rho_\text{max}$ and the elastic energy are
\begin{eqnarray}
  \rho_\text{max} &=& \frac{8}{3\pi} \sqrt{s_c/2} \, \sqrt{\Delta x} \approx 1.026\, \sqrt{\Delta x} \; ,  \qquad \text{and}
  \label{eq:rhomaxlin}\\
  e &=& \int_{-1}^1 \frac{s_c}{4} \left( \psi'^2 + 12 \rho^2 \right) \romd \tilde{s} = \left( \frac{5\pi^2}{ s_c^2} + \frac{20 s_c^2}{ 3\pi^2} \right) \, \Delta x \approx f^{(0)} \, \Delta x 
  \; .
\end{eqnarray}
We finally discuss the existence of periodic solutions consisting of multiple identical folds of equal length. As before, each fold is symmetric (asymmetric in $\psi(\tilde{s})$) and its boundary conditions are $\psi^{(1)}(-1)=\psi^{(1)}(+1)=0$. Eq.~(\ref{eq:COND0}) thus remains valid. The curvature at the boundary, however, is not  zero anymore for periodic solutions (periodic solutions with $c_1=0$ are unstable, as we will observe numerically in the next section). Denoting its value at $\tilde{s}=1$ with $c_1$, the curvature boundary condition becomes
\begin{eqnarray}\label{eq:CONDC0}
 \left[\sqrt{ \left|  \lambda_-^2 \right| } \cos \left( \sqrt{ \left|  \lambda_-^2 \right| }\right) -  \sqrt{\left| \lambda_+^2 \right| } \sin \left( \sqrt{ \left| \lambda_-^2 \right|} \right) 
 \cot \left( \sqrt{ \left| \lambda_+^2 \right| }\right) 
 \right] - \frac{c_1}{a_2} = 0
 \; .
\end{eqnarray}
To conserve the Hamiltonian, Eq (\ref{eq:rescaledHamiltonian}) now yields the condition
\begin{equation}\label{eq:CONDC1}
\frac{2 ( \psi'(0)^2 - c_1^2 )}{s_c^2} - \frac{3 s_c^2}{2} \left( \int_{-1}^0 \sin \psi \, \romd \tilde{s}\right)^2 = 0
\; .
\end{equation}
Equations (\ref{eq:CONDC0}) and (\ref{eq:CONDC1}) leave the ratio $c_1/a_2$ of amplitude $a_2$ and curvature $c_1$ undetermined. We now proceed as follows: Since $c_1>0$ for radially outward buckling, and $a_2<0$, we prescribe a negative ratio $c_1/a_2$ and search for roots $(\tilde{f}^{(0)}, s_c^{(0)})$ of Eqns.~(\ref{eq:CONDC0}, \ref{eq:CONDC1}). 
Since we are only interested in the mode with lowest buckling force, we numerically determine roots of (\ref{eq:CONDC0}) and (\ref{eq:CONDC1}) in the vicinity of the zero-curvature solution, Eq.~(\ref{eq:lowestsolution}). The inset of Fig.~\ref{fig:linearization} shows that the minimal buckling force $f^{(0)}$ increases for $c_1/a_2<0$, and decreases for positive $c_1/a_2$ (dotted line). Since positive values of $c_1/a_2$ correspond to unphysical inward buckling, the inset suggests that the mode with lowest buckling force is the single fold with $c_1/a_2=0$, and periodic solutions are not selected at the buckling onset. This result is in contrast to the periodic undulations found for an axially compressed cylinder in absence of a  substrate \cite{Timoshenko59}.

We can easily verify that the zero-curvature solutions with $c_1=0$ is a saddle point (or local extremum) of the buckling force by considering small perturbations around $c_1=0$: Equations (\ref{eq:CONDC0}) and (\ref{eq:CONDC1}) geometrically describe two surfaces $\mathcal{S}_1$ and $\mathcal{S}_2$ in the space of independent parameters $(f^{(0)},s_c^{(0)}, c_1/a_2)$. Both equations are simultaneously satisfied along the curve of intersection $\gamma$ of $\mathcal{S}_1$ and $\mathcal{S}_2$. Denoting the parameters of the single fold solution as $p_* = (20/\sqrt{3}, \left(3/4 \right)^{1/4} \pi, 0)$, the tangent $\vec{T}|_{p_*}$ to $\gamma$ at $p_*$ is given by the line of intersection between the two tangent planes of $\mathcal{S}_1$ and $\mathcal{S}_2$. Basic geometry then yields 
\begin{equation}\label{eq:intersectiongradient}
\vec{T}|_{p_*} = \nabla \mathcal{S}_1|_{p_*} \times \nabla \mathcal{S}_2|_{p_*} = - \left( 
0, \frac{1}{4} (3 \pi ), \frac{3^{3/4} \pi ^2}{\sqrt{2}}  \right)
\; ,
\end{equation}
where $\nabla \mathcal{S}_{1,2}|_{p_*}$ are the gradients of $\mathcal{S}_{1,2}$ evaluated at $p_*$ with respect to the parameters $(f^{(0)}, s_c^{(0)}, c_1/a_2)$. Eq. (\ref{eq:intersectiongradient}) shows that the buckling force does not change near $p_*$, which agrees with $f^{(0)}$ having a saddle-point (or local extremum) at $p_*$.


\section{Numerical solutions and experiments\label{sec:numericalsolutionsexperiments}}

\begin{figure}[t]
  \begin{center}
  \subfigure[][]{\label{fig:EnergyandFvsDeltax}\includegraphics[height=0.45\textwidth]{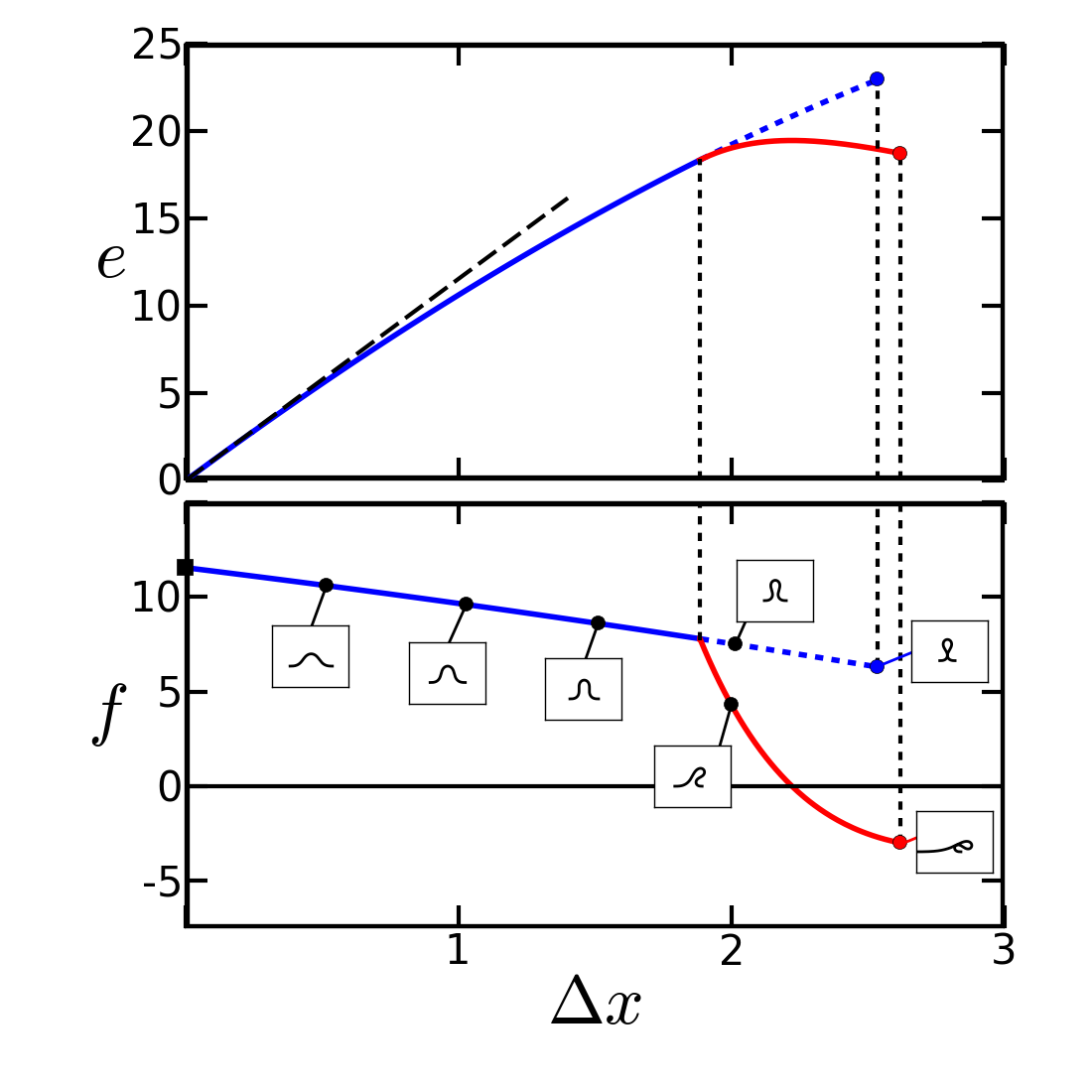}}
  \hfill
  \subfigure[][]{\label{fig:FnvsFx}\includegraphics[height=0.45\textwidth]{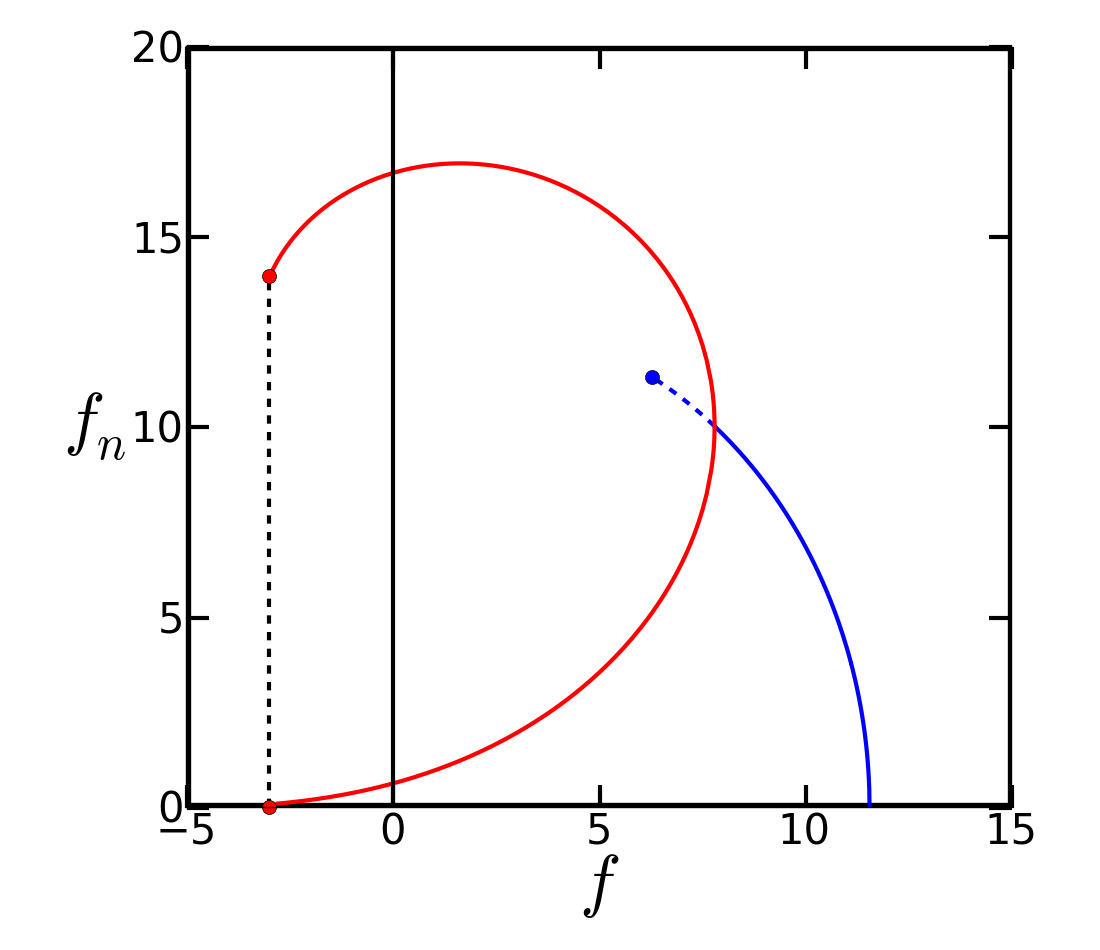} }
  \caption{\label{fig:RungeKuttaresults}(a)  Total elastic energy $e$ \textit{(top)} and horizontal force $f$ \textit{(bottom)} as a function of the compressive axial displacement $\Delta x$ for the scaled nonlinear system (Eqns.~(\ref{eq:Hamiltonequations})/(\ref{eq:boundaryconditions})). For small compressions the solutions are symmetric (blue solid branch) and coincide with the solutions of the linearized system (dashed black line \textit{(top)} and black square \textit{(bottom)}). Below a compressive force of approx.\ 7.8, asymmetric solutions appear (red branch) which are energetically favoured over the symmetric ones (blue dotted branch). The two branches are depicted up to the point of self-contact (assuming a vanishing thickness $h=0$ of the sheet). \textit{Insets}: Profiles for different values of $\Delta x$.
  (b) Normal force at the circular boundaries, $f_n$, as a function of the horizontal force $f$ for the scaled nonlinear system (Eqns.~(\ref{eq:Hamiltonequations})/(\ref{eq:boundaryconditions})).  Its value is the same on both boundaries when the fold is symmetric. For the antisymmetric profile two different values are found when fixing $f$. The higher one corresponds to the boundary to which the fold has tipped over. }
  \end{center}
\end{figure}
To find the shape of the sheet for high deformations we solve the Hamilton equations~(\ref{eq:Hamiltonequations}) numerically 
using a standard shooting method \cite{NumRec}: for a fixed scaled force $f$ and a trial value for 
$p_\rho=-f_n$ at $s=0$ the equations are integrated with a fourth-order Runge Kutta method. 
For any trial $f_n$, the values of $\psi$, $\rho$, and $p_{\psi}$ at $s=0$ are obtained from the 
boundary conditions~(\ref{eq:boundaryconditions}). The integration is stopped as soon as $\rho=0$ is reached again. 
Every $f_n$ which results in $\psi (s_c) = 0$ corresponds to the profile of a single fold (see Fig.~3).

For low values of compression ($\Delta x\ll 1$) the nonlinear solutions are symmetric and coincide with the ones of the linear regime as expected (see Fig.~\ref{fig:EnergyandFvsDeltax}). The associated compressive force $f$ decreases with $\Delta x$. The negative sign of $\frac{\partial f}{\partial \Delta x}$ implies that solutions with more than one fold are unstable: in equilibrium the force $f$ has to be a constant along the whole profile for a fixed $\Delta x$. If the profile consisted of $N$ folds, each fold $i$ would correspond to a solution of the nonlinear system with $(\Delta x)_i=\Delta x / N \, (i\in\{1,\dots,N\})$. A small compressive fluctuation $(\Delta x)_i \to (\Delta x)_i + \delta$ would decrease the force $f_i$ needed to stabilize the fold $i$. Since the sheet is unstretchable in the longitudinal direction, the adjacent fold would be less compressed than before. To keep it in place, however, a force which is higher than $f$ and $f_i$ would be needed. Since this is not possible, the entire system becomes unstable. In particular, periodic solutions are unstable irrespective of the curvature at the boundary. Thus, unlike most other setups involving compressed sheets on deformable substrates \cite{Pocivavsek08,Brau13,Brau11,DiamantWitten13,Santangelo14}, our system does not exhibit a transition from periodic wrinkles to localized folds; instead, it forms a single fold as soon as $\Delta x\neq 0$.

Above a critical axial displacement $\Delta x_\text{crit} \approx 1.88$ the symmetry of the solutions is broken: the global energy minimum corresponds to an asymmetric fold. The associated force $f$ decreases quicker than before and changes sign at $\Delta x\approx 2.22$. At this displacement value the energy displays a maximum and the corresponding external force is zero. 
For even higher $\Delta x$ we obtain a recumbent fold and one has to pull instead of to compress to hold the asymmetric sheet in equilibrium. When the sheet is confined further it starts to touch itself. Assuming a vanishing thickness $h$, one finds $\Delta x\approx 2.62$ for the asymmetric and $\Delta x\approx 2.54$ for the symmetric branch. 


\begin{figure}[t]
  \begin{center}
   \subfigure[][]{\label{fig:DeltaRvsDeltax}\includegraphics[width=0.49\textwidth]{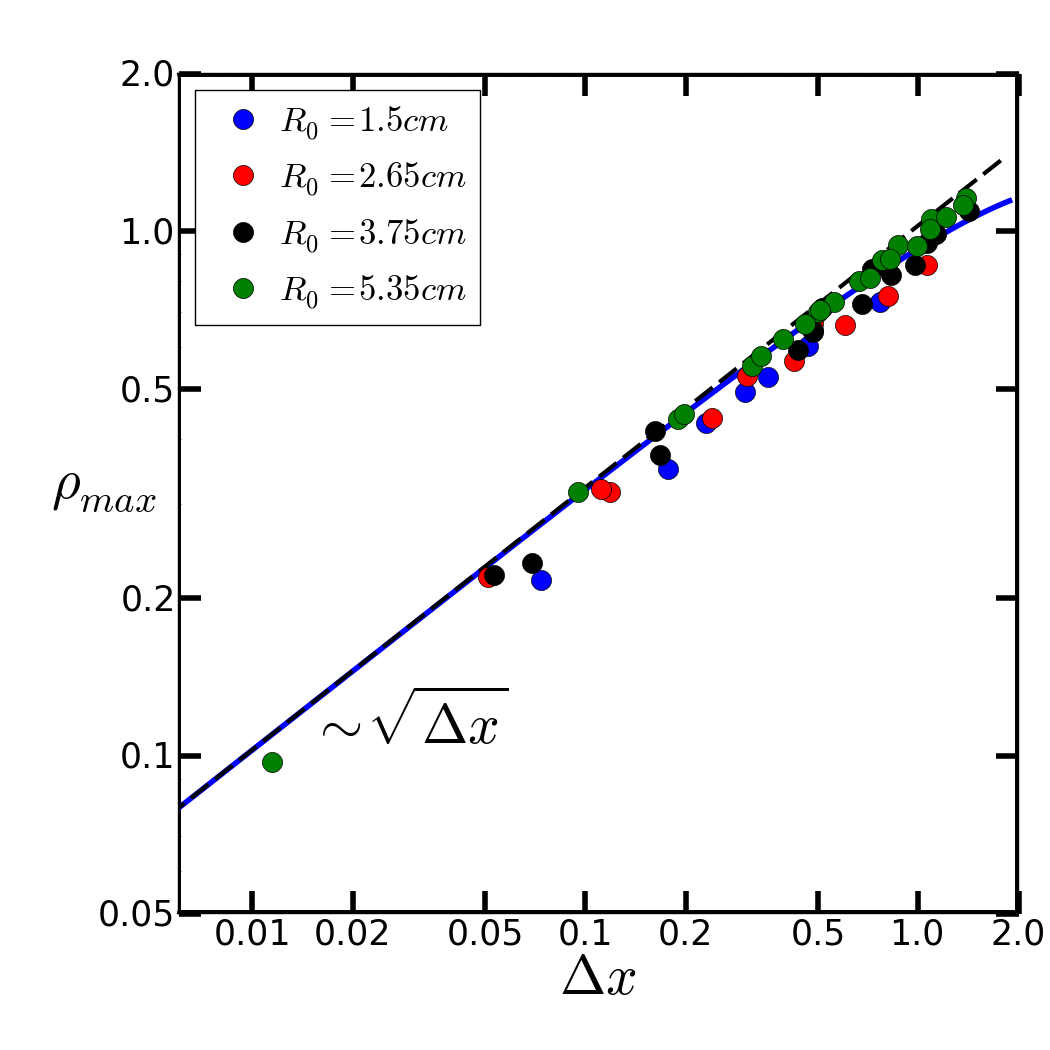}}
   \hfill 
   \subfigure[][]{\label{fig:profiles}\includegraphics[width=0.49\textwidth]{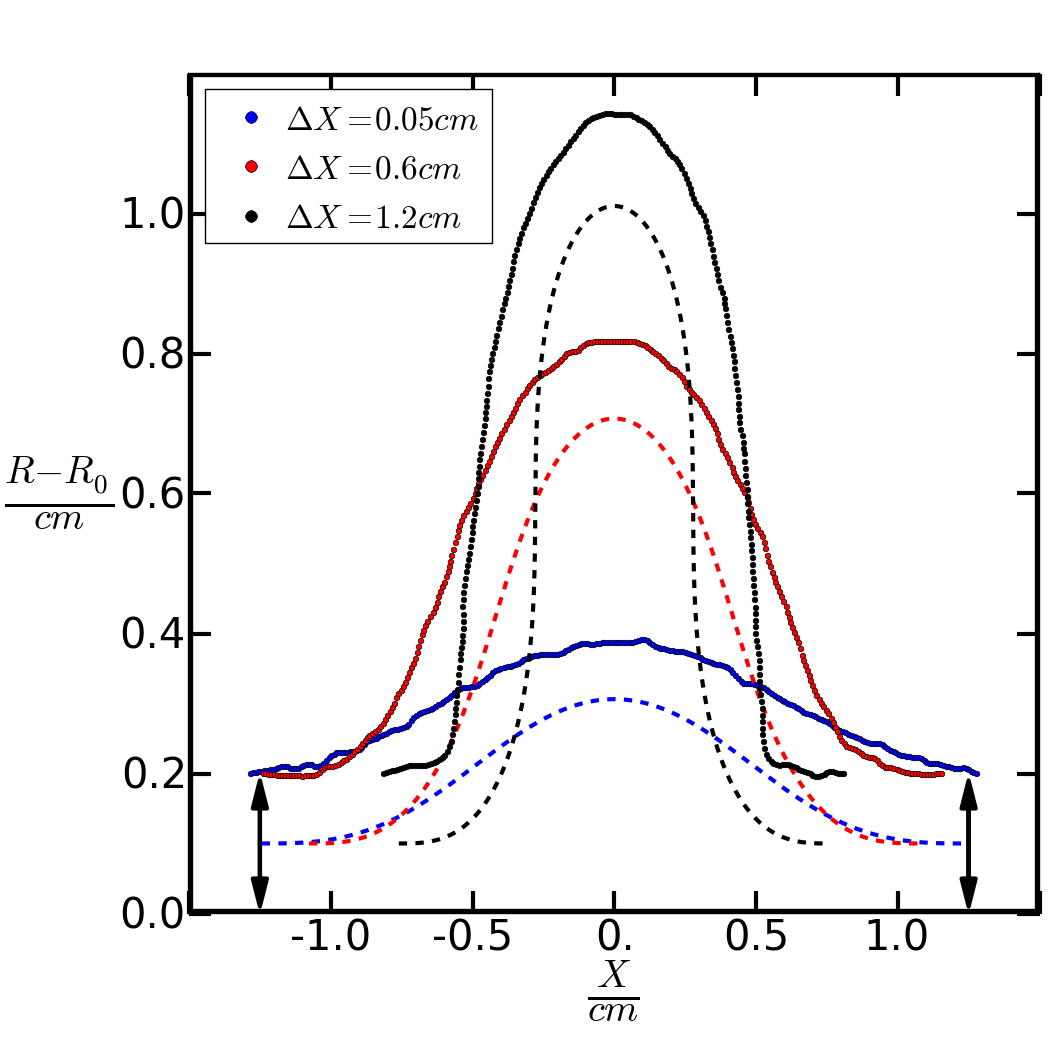}}
       \caption{\label{fig:experiments} (a) Maximum radial displacement $\rho_\text{max}$ as a function of $\Delta x$ for a neoprene sheet jackating a cylinder of radius $R_0$ (dots). To compare the results to the nonlinear solution $\rho(s = s_c / 2)$ of the scaled system (solid blue line), all measurements were divided by $\sqrt{0.2\text{cm}\, R_0}$.  The black dashed line shows the solution in the linear regime (Eq.~(\ref{eq:rhomaxlin})).
       (b) Comparison between experimental (points) and theoretical nonlinear profiles (dotted lines) for $R_0=3.75$cm and different values of $\Delta X$. All profiles are plotted in unscaled units. The thickness $h=0.2$cm of the sheet is indicated with double arrows.}
  \end{center}
\end{figure}
We validate our theoretical results $(i)$ with the help of finite element simulations (see appendix), and $(ii)$ in a series of experiments with neoprene sheets wrapped around cylinders of different radii. In these experiments, the sheet is prepared in such a way that it is free of stretching. The experimental profiles are then photographed and extracted with ImageJ \cite{ImageJ}. 
While relatively simple, we note that this technique only allows the analysis of symmetric profiles before the tipping point.\footnote{As soon as the profile becomes asymmetric one cannot infer the exact profile any more because the part below the fold is hidden from the camera.}
Fig.~\ref{fig:experiments} shows the experimental results together with the theoretical predictions. Since all material parameters scale out of the theory, no fitting parameters are needed. In Fig.~\ref{fig:DeltaRvsDeltax} the maximum value of the radial displacement, $\rho_\text{max}$, is plotted as a function of the axial compression $\Delta x$. One observes that the experimental curves converge towards the theoretical solution for increasing radius $R_0$. This can be explained by taking a look at our theoretical assumptions again: to simplify the problem we have assumed that $R_0/h$ is large (see Sec.~\ref{sec:model}). 
The thickness of the neoprene sheet is constant and approximately $h=0.2$cm in all experiments, implying that our theoretical approximations become more accurate for larger radii $R_0$. The effect of the cylinder radius is also noticeable when looking at the tipping point: while theoretically predicted to occur at $\Delta x \approx 1.88$, experiments yield smaller values that decrease with $R_0$. 

Fig.~\ref{fig:profiles} shows three different experimental profiles together with the corresponding nonlinear theoretical solutions for a cylinder radius of $R_0=3.75$cm. Note that experimental and theoretical results are shifted radially, since the digitized profiles describe the outer side of the sheet ($h=0.2$cm away from the solid cylinder) whereas the nonlinear solutions correspond to the theoretical centerline of the sheet. Despite the simplified nature of our model, theory and experiment coincide remarkably well even for high curvatures at the tip of the fold.


\section{Conclusion\label{sec:conclusion}}
Euler buckling is a well-studied phenomenon with numerous applications and occurs in various situations in nature and engineering \cite{AudolyPomeau}. In its simplest form it deals with the situation of translatory invariance in one direction, such as the buckling on planar substrates  \cite{Pocivavsek08}. However, buckling often occurs in curved geometries \cite{Chen10,Li12} inducing non-isometric deformations of the sheet. In this article we have exemplified the effect of curvature on wrinkling by considering an Euler-type buckling in a cylindrical geometry. In our setup, an elastic membrane is wrapped around a cylindrical substrate of same radius and compressed axially. Instead of a wrinkle-to-fold transition typical for elastic substrates, a single fold appears immediately above the buckling threshold. For even larger compression, the fold breaks symmetry, leading eventually to negative buckling forces. The theoretical predictions of our simple model were validated with experiments on neoprene sheets and finite element simulations. Without a fitting parameter, experiments, simulations and theory coincide rather remarkably. 
 
While we do not observe wrinkling solutions,  we speculate that such periodic solutions would form only if the radius of the \textit{unstretched} elastic sheet was larger than $R_0$: For small amplitudes, the sheet would not be in contact with the cylindrical substrate and thus form the periodic undulations of a free, axially compressed cylindrical sheet. For larger compression, contact with the rigid substrate eventually occurs, which then most likely leads to the single fold solutions discussed in this article.

In the experiments one can force the sheet over the point of self-contact. Due to the friction between the sheet and the substrate a stable equilibrium can be found whose $\Delta x$ depends on the value of the friction coefficient. Our system can thus be interpreted as a simple mechanical switch with two stable states: the cylindrical geometry and an asymmetric configuration with $\Delta x > 2.62$. To switch from the cylindrical state to the other, an external compressive force above the critical buckling force $f=11.55$ has to be applied. To return to the unstretched sheet requires stretching the fold (decreasing $\Delta x$) into a regime of positive $f>0$, after which it will unfold on its own. Consisting only of a single movable part and featuring simple assembly steps, such a switch could potentially be used in micro- and nanoscale  applications.


\begin{acknowledgments}
This work was supported by the Swiss National Science Foundation grant No. 148743 (N.~S.). 
The authors thank Romain Lagrange for fruitful discussions.
\end{acknowledgments}


\appendix
\section{}

\subsection{Derivation of the elastic energy}

We derive the elastic energy of the cylindrical shell by first linearizing the elastic strains associated with a deformation of the cylinder. Comparing the deformed and undeformed infinitesimal length of two neighbouring material points, one finds for the stretching strains at linear order in $R$:
\begin{eqnarray}
\epsilon_{SS} &=& 0 \; ,\\
\epsilon_{\varphi \varphi} &=& \frac{R - R_0}{R_0} \label{eq:strain1}
\; ,
\end{eqnarray}
where $\varphi$ measures the angle in the circumferential direction on the cylinder. Note that the first equation follows from the inextensibility in the $S$-direction (see next section), and the off-diagonal $\epsilon_{S \varphi} = 0$ for axisymmetric twistless shells. Similarly one obtains for the bending strains  
\begin{eqnarray}
K_{SS} &=& \psi' \; ,\\
K_{\varphi \varphi} &=& \kappa_{\varphi \varphi} - \kappa^0_{\varphi \varphi} = \frac{\cos \psi}{R} - \frac{1}{R_0}
 \; ,
\end{eqnarray}
where $\kappa_{\varphi \varphi}$ and $\kappa^0_{\varphi \varphi}$ are the curvatures in the circumferential direction of the deformed and undeformed configuration, respectively. Due to axisymmetry, the off-diagonal strain again vanishes,  $K_{S \varphi}=0$. Using Eq.~(\ref{eq:strain1}), we write 
$K_{\varphi \varphi} = \frac{1}{1+\epsilon_{\varphi \varphi}} \frac{\cos \psi}{R_0} - \frac{1}{R_0}\;.$
Since the bending strains will only dominate for deformations close to isometric ones, where the stretching strains are $\approx 0$, we may further approximate $K_{\varphi \varphi} \simeq \frac{\cos \psi}{R_0} - \frac{1}{R_0}$ \cite{AudolyPomeau}, leading to the final expressions for the linearized strains
\begin{eqnarray}\label{eq:finalstrains}
\epsilon_{\varphi \varphi} &=& \frac{R-R_0}{R_0} \; ,\\
K_{SS} &=& \psi' \; ,\\
K_{\varphi \varphi} &=& \frac{\cos \psi}{R_0} - \frac{1}{R_0} \; ,\\
\kappa_{S\varphi} &=& \epsilon_{SS} = \epsilon_{S \varphi} = 0
 \; .
\end{eqnarray}
Using an isotropic Hookean material with Young's modulus $Y$ and Poisson ratio $\nu$, the simplest form of energy densities with decoupled bending and stretching contributions reads \cite{AudolyPomeau}
\begin{eqnarray}
\Phi &=& \frac{Yh}{2(1-\nu^2)} \left( \epsilon_{SS}^2 + \epsilon_{\varphi \varphi}^2 + 2 \nu \epsilon_{SS} \epsilon_{\varphi \varphi}\right) \; ,\\
\Psi &=& \frac{Yh^3}{24(1-\nu^2)} \left(  K_{SS}^2 + K_{\varphi \varphi}^2 + 2 \nu K_{SS} K_{\varphi \varphi} \right)
 \; .
\end{eqnarray}
The total energy is then obtained by integration over the undeformed reference surface, 
\begin{equation}
E = \int_{0}^{S_c} \int_0^{2\pi} \Phi(\epsilon_{SS},\epsilon_{\varphi \varphi}) + \Psi(K_{SS},K_{\varphi \varphi}) R_0 \, \romd \varphi \, \romd S 
\end{equation}
from which Eq.~(\ref{eq:energy1}) is obtained.


\subsection{Comparison with finite element simulations}

\begin{figure}[t]
  \begin{center}
   \subfigure[][]{\label{fig:abaqusprofile}\includegraphics[width=0.42\textwidth]{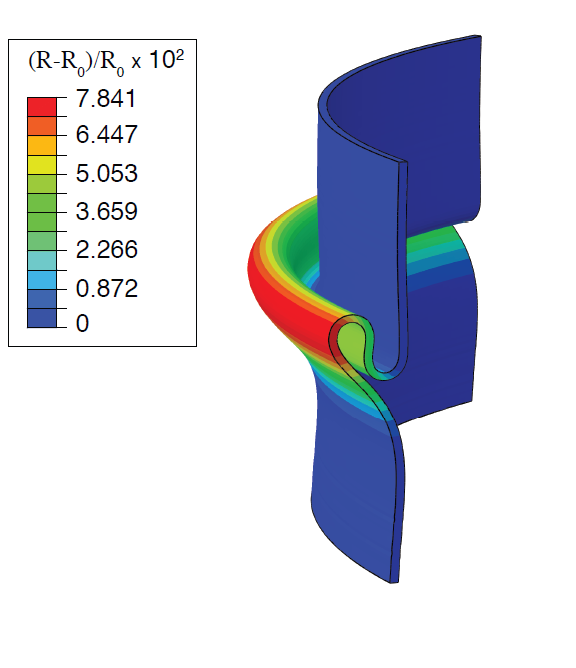}}
   \hfill
   \subfigure[][]{\label{fig:abaquscomparison}\includegraphics[width=0.49\textwidth]{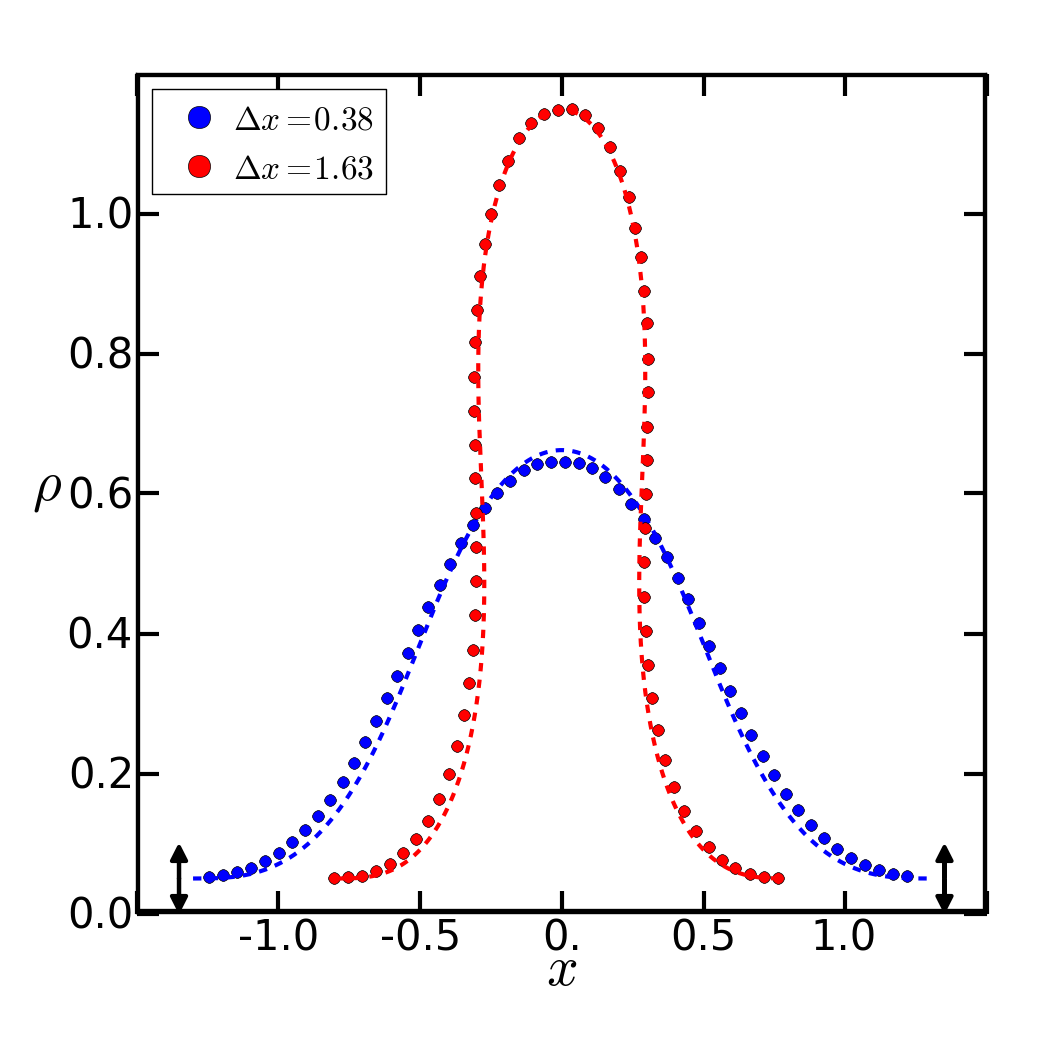}}
       \caption{\label{fig:abaqus} (a) \textsc{Abaqus} simulation of a recumbent fold for $R_0=1.0$, $h=0.01$ and $\Delta X=0.232$. Color-coded is the relative radial displacement $(R-R_0)/R_0$. Only a segment of the cylindrical membrane is shown for better clarity.
       (b) Comparison between \textsc{Abaqus} profiles (points) and theoretical nonlinear profiles (dotted lines) for two values of $\Delta x$. All profiles are plotted in scaled units. The thickness of the sheet is indicated with double arrows.}
  \end{center}
\end{figure}

The mechanical model leading to Eq.~(\ref{eq:energysimplemodel}) assumes linearity of strains, inextensibility along the axial direction, and radii of curvatures that are large with respect to the shell thickness $h$. To qualitatively test the validity of these assumptions, we compare the deformed membrane profiles of Eq.~(\ref{eq:energysimplemodel}) with numerical simulations obtained from the commercially available \textsc{Abaqus} finite element software package \cite{abaqus}. In \textsc{Abaqus}, the  membrane is modeled as a three-dimensional body using the reduced-integration quadrilateral elements CAX4R, which take the axisymmetry of the problem into account to reduce computational cost. We found that a spatial resolution of 2 elements in the thickness direction is sufficient to capture the deformation accurately, with higher resolutions leading to no observable change in the profile geometry. The rigid cylindrical constraint is modeled using Hertzian contact dynamics, penalizing finite element nodes that impenetrate the cylinder \cite{Popov2010}. Since the fold length $S_c$ changes during axial compression but is not known \textit{a priori}, we simulated a cylindrical membrane of length $L \gg S_c$. The geometries considered are $h/R_0$ between $5\cdot 10^{-4}$ and $10^{-2}$. 
At the membrane boundary, $R(0)=R(L)=R_0$ and $\psi (0)  = \psi (L) = 0$ is imposed. To compress the membrane, we 
prescribe the axial displacement $U_x$ by enforcing $U_x(0)=0$ and $U_x(L)=-\Delta X$. Starting from $\Delta X=0$, we slowly increase $\Delta X$ to obtain fold profiles at various stages of compression. 
\par
Fig.~\ref{fig:abaqusprofile} shows an axisymmetric fold obtained from a typical simulation (for better clarity, only a part of the membrane is shown), agreeing qualitatively with the recumbent fold found in experiments and using the simplified model~(\ref{eq:energysimplemodel}). To investigate the validity of axial inextensibility, we compare  profiles in the early and intermediate stages of folding, where errors introduced by inextensibility are expected to be maximal due to the large compressive forces present in the fold. Fig.~\ref{fig:abaquscomparison} shows that our simplified model compares well to the profiles of \textsc{Abaqus} even for large compressive forces, demonstrating, in particular, that inextensibility is a valid model assumption.


\end{document}